\documentclass[%
twocolumn,
reprint,
amsmath,amssymb,
aps,
prl
]{revtex4-2}

\usepackage{graphicx}
\usepackage{dcolumn}
\usepackage{bm}
\usepackage{hyperref}
\usepackage[normalem]{ulem} 

\hypersetup{
	colorlinks=true,
	linkcolor=blue,
	filecolor=magenta,      
	urlcolor=blue,
	citecolor=blue,
	anchorcolor=blue,
}

\begin{document}
	

\title{Trimerized Spin-$1/2$ Chain: Emergent Low-Energy Hamiltonian, Higher-Energy Excitations, and Magnetic and Thermodynamic Responses }
	
	\author{Snehasish Sen}
	
	\author{Sudhansu S. Mandal}
	
	\affiliation{
		Department of Physics, Indian Institute of Technology, Kharagpur, West Bengal 721302, India
	}%
	
	%
	%
	
	\date{\today}

\begin{abstract}
 We investigate a spin-1/2 antiferromagnetic chain with trimerized exchange couplings relevant to Na$_2$Cu$_3$Ge$_4$O$_{12}$. We show that the low-energy Hilbert space maps onto an effective Heisenberg chain of composite spin-1/2 trimer degrees of freedom with positive coupling, enabling an exact spinon excitations via the Bethe ansatz. To access higher-energy dynamics, we develop a self-consistent Jordan-Wigner mean-field theory that yields three fermionic bands reflecting the underlying trimer structure. Remarkably, this approach reproduces the exact low-energy spinon continuum while predicting two additional higher-energy excitation bands consistent with the experimental observations and previous numerical simulations. The theory further captures the 1/3 magnetization plateau under an applied magnetic field, and provides testable predictions in magnetic susceptibility and specific heat. Our results establish a unified framework connecting low- and high-energy excitations in trimerized quantum spin chains.
\end{abstract}

\maketitle

Low-dimensional quantum spin systems with nonuniform exchange interactions\cite{KITAEV2006,Nigam2025} provide a fertile platform for realizing exotic excitations and emergent quasiparticles. In particular, spin-1/2 chains with periodically varying couplings\cite{Bulaevskii1963,Duffy1968,Chitra1995,Brenig1997,Wang2013} can host composite degrees of freedom and nontrivial excitation spectra beyond those of the uniform Heisenberg chain. Recent experiments \cite{Bera2022} on the trimerized compound Na$_2$Cu$_3$Ge$_4$O$_{12}$ (NCGO) have revealed rich magnetic behavior, including a 1/3 magnetization plateau and multiple excitation bands in inelastic neutron scattering, indicating the presence of correlated dynamics beyond conventional spinon descriptions.

While the low-energy physics of uniform\cite{Cloizeaux1962,Yamada1969,Tennant1995,Zaliznyak2004,Lake2005,Lake2010,Mourigal2013} and dimerized Heisenberg chains is well understood -- through exact solutions such as the Bethe Ansatz\cite{Bethe1931,Hulthen1938} and models such as the Majumder-Ghosh chain\cite{Majumdar1969}--the nature of excitations in trimerized spin systems remains less explored. In particular, it is unclear how the interplay between intra-trimer and inter-trimer couplings give rise to both gap-less low-energy excitations and additional higher-energy modes observed experimentally. A unified theoretical framework that captures these features within a minimal microscopic model is still lacking. 

In this work, we study a spin-1/2 antiferromagnetic chain with periodically modulated exchange couplings of the form $J_1$--$J_1$--$J_2$, $(J_1>J_2)$, which realizes a natural trimer structure and serves as a minimal model for NCGO. We show that the low-energy Hilbert space is spanned by the effective spin-1/2 degrees of freedom associated with trimer ground states, leading to an emergent Heisenberg chain with renormalized coupling $4J_2/9$. This mapping allows an exact characterization of the low-energy sector in terms of spinon excitations obtained from the Bethe Ansatz.

To go beyond the low-energy description, we employ a Jordan-Wigner transformation\cite{Jordan1928} followed by a self-consistent mean-field analysis, which yields three fermion bands corresponding to the underlying trimer structure, and provides the description of the ground state as fully filled lowest band and the half-filled middle band. We demonstrate that the excitations within this framework not only reproduce the exact spinon continuum, but also give rise to two additional higher-energy bands, consistent with the numerical\cite{Cheng2022,Bera2022,Cheng2024,Li2025,Prabhakar2025} and experimental observations\cite{Bera2022}. Furthermore, the theory captures the emergence of the 1/3 magnetization plateau under an applied magnetic field, and predicts low- and high-temperature characteristics of magnetic susceptibility and specific heat. 

Our results provide a unified description of low- and high-energy excitations in trimerized quantum spin chains and establish a direct connection between microscopic interactions, emergent quasiparticles, and experimentally observable response functions.

	\begin{figure}[h] 
		\centering   
		\includegraphics[width = \linewidth]{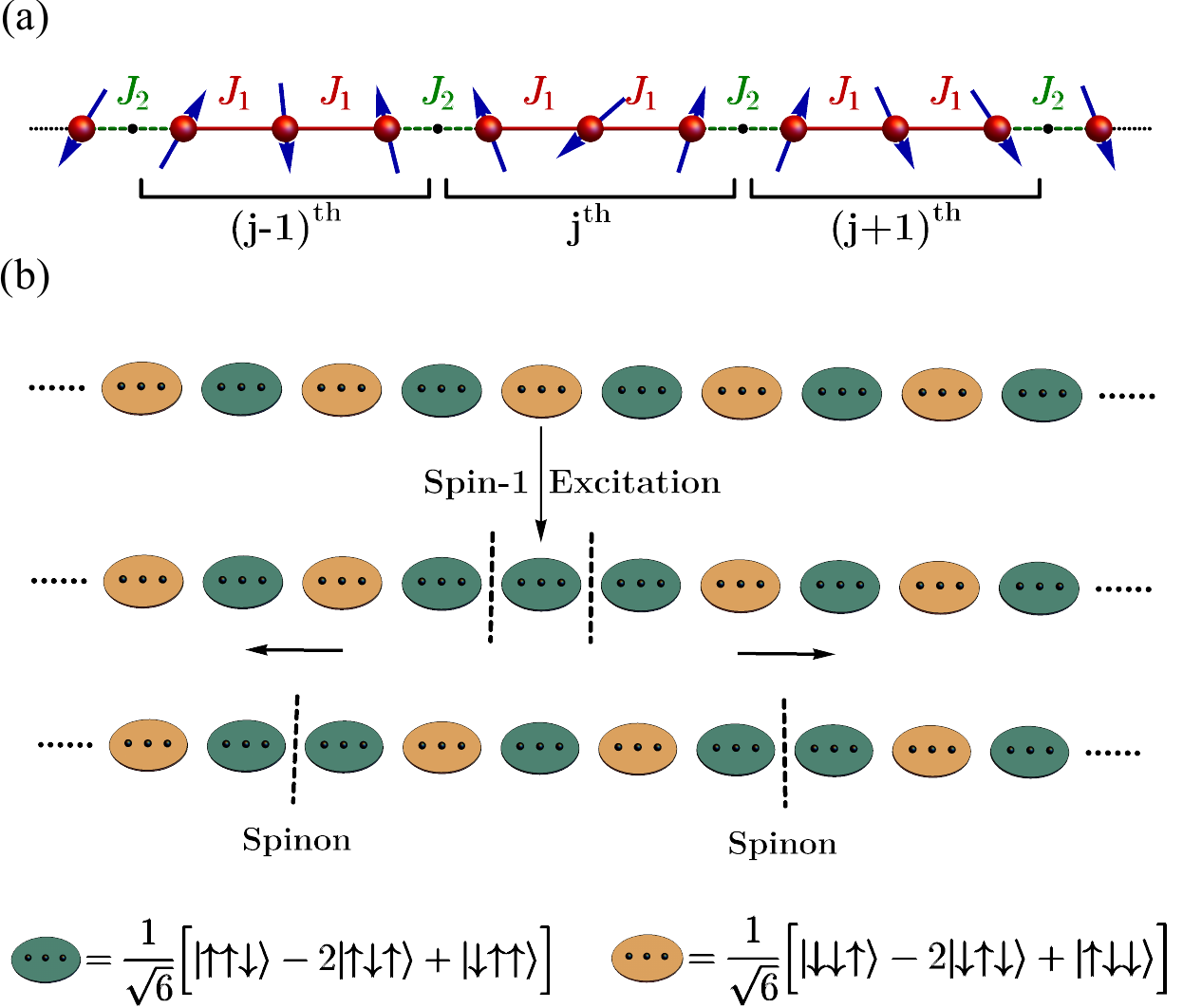}
		\caption{(a) A schematic of a spin chain of spin-1/2. Bonds with  Heisenberg exchange interaction strengths are shown. Trimer unit cells: $(j-1)$-th, $j$-th,  and $j+1$-th are marked. (b) Composite spin-1/2 doublets of trimers are denoted by \raisebox{-0.3ex}{\includegraphics[height=1em]{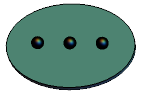}} and \raisebox{-0.3ex}{\includegraphics[height=1em]{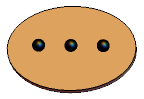}}.
			Top: Schematic of antiferromagnetically correlated ground state. Middle: Schematic of spin-1 excitation bounded by two domain walls. Bottom: Domain walls shift with time as representation of a pair of spinon excitations. }
		\label{schematic}
	\end{figure}

 A minimal spin model of the NCGO compounds \cite{Bera2022,Mo2006,Yasui2014}, 
  may be elucidated schematically in Fig.~\ref{schematic}a in terms a spin-1/2 chain with  the periodically modulated antiferromagnetic couplings of $J_1$--$J_1$--$J_2$ with $J_1>0$, $J_2 >0$, and $\gamma = J_2/J_1 <1$.  $\gamma \neq 1$ makes the system unique having a periodic arrangement of spin trimers with three sub-lattices, while $\gamma =1$ corresponds to Heisenberg spin chain having pioneering exact Bethe-Ansatz solution. In Ref.~\onlinecite{Bera2022}, it is pointed out that there is a second nearest neighbor interaction within a trimer only, but no inter-trimer second nearest neighbor interaction. This merely shifts the energy eigenvalues of a single trimer. Ignoring this intra-trimer interaction does not alter any qualitative physics of the system. 
 The Hamiltonian for the chain consisting of $N$ trimers is thus considered for simplicity as, 
 {\small  \begin{eqnarray}\label{full_hamiltonian}
  {\cal H} &=& \sum_{j = 1}^{N}\Big[ J_1 \left(\vec{S}_{j,1}\cdot\vec{S}_{j,2}+\vec{S}_{j,2}\cdot\vec{S}_{j,3}\right) 
  + J_2 \vec{S}_{j,3}\cdot\vec{S}_{j+1,1}\Big] 
  \end{eqnarray}
}where $\vec{S}_{j,\lambda}$ $(\lambda =1,2,3)$ represents spin-1/2 operators of $j$-th trimer and $\lambda$-th sub-lattice. Here, the first term indicates  the Hamiltonian of a single trimer ${\cal H}_j $ and the second term indicates the Hamiltonian ${\cal H}_{j,j+1}$ between two consecutive trimers. ${\cal H}_j $ can be solved exactly: First, a doubly degenerate ground state with energy $\epsilon_0 = -J_1$ having eigen states $\vert +\rangle = \frac{1}{\sqrt{6}}\left( \vert \uparrow\uparrow \downarrow\rangle - 2 \vert\uparrow \downarrow\uparrow\rangle +\vert \downarrow\uparrow \uparrow \rangle \right)$ and $\vert -\rangle = \frac{1}{\sqrt{6}}\left( \vert \downarrow\downarrow \uparrow\rangle - 2 \vert \downarrow \uparrow\downarrow\rangle + \vert\uparrow\downarrow \downarrow \rangle \right)$ in terms of spin configurations in the consecutive spins of the trimer is obtained, where $\vert \uparrow,\downarrow\rangle$ are the eigenstates of the Pauli spin-operator $\sigma_z$ corresponding to a single spin. These $\vert \pm \rangle$ are eigen states of another Pauli spin matrix $\tau_z$ corresponding to spin-1/2 operator $\vec{d}=\frac{1}{2} \vec{\tau}$ of the resultant spin of the trimer. Second, the first excited state with energy $\epsilon_1 = 0$ is also doubly degenerate having eigenstates $ \frac{1}{\sqrt{2}}\left( \vert \uparrow\uparrow \downarrow\rangle -\vert \downarrow\uparrow \uparrow \rangle \right)$ and $\frac{1}{\sqrt{2}}\left( \vert \downarrow\downarrow \uparrow\rangle - \vert\uparrow\downarrow \downarrow \rangle \right)$ corresponding to another spin-1/2 configuration of the trimer, denoted by $\vec{D}$. Third, the quadruply degenerate eigen states $\vert \uparrow \uparrow \uparrow\rangle$, $\vert \downarrow\downarrow\downarrow \rangle$, $\frac{1}{\sqrt{3}}\left( \vert \uparrow\uparrow \downarrow\rangle +  \vert\uparrow \downarrow\uparrow\rangle +\vert \downarrow\uparrow \uparrow \rangle \right)$, and $ \frac{1}{\sqrt{3}}\left( \vert \downarrow\downarrow \uparrow\rangle + \vert \downarrow \uparrow\downarrow\rangle + \vert\uparrow\downarrow \downarrow \rangle \right)$ corresponding to total spin 3/2 of the trimer, denoted by $\vec{Q}$, having energy $\epsilon_2 = J_1/2$ is obtained.

The matrix elements of inter-trimer Hamiltonian ${\cal H}_{j,j+1}$ in the low-energy Hilbert space spanned by spin-1/2 states of two consecutive trimers $\vert +\rangle_j \,\vert +      \rangle_{j+1}$, $\vert +\rangle_j \,\vert -      \rangle_{j+1}$, $\vert -\rangle_j \,\vert +      \rangle_{j+1}$, and  $\vert -\rangle_j \,\vert -      \rangle_{j+1}$ are found as
\begin{equation}
	J_2 \left( \begin{array}{rrrr}  \frac{1}{9} & 0 & 0 & 0\\
		0 & -\frac{1}{9} & \frac{2}{9} & 0 \\
		0 & \frac{2}{9} & -\frac{1}{9} & 0 \\
		0 & 0 & 0 & \frac{1}{9} \end{array} \right)
	\label{matrix-4}
\end{equation}
which is nothing but the Hamiltonian of Heisenberg interaction between two spin-1/2 objects with antiferromagnetic coupling constant $4J_2/9$.  Also, no interaction is present (see end matter) between next-nearest neighbor trimers having  {\em low-energy} Hilbert space.             	
We thus obtain an effective Hamiltonian in the {\em low-energy} Hilbert space of $N$ trimers as
\begin{equation}
	{\cal H}_{\rm eff} = N\epsilon_0 + \frac{4}{9} J_2 \sum_{j=1}^N \vec{d}_j \cdot \vec{d}_{j+1}
	\label{H_effective}
\end{equation}
which signifies an antiferromagnetic chain of spin-1/2 trimer spins. Since Eq.~\eqref{matrix-4} exactly represents bilinear Heisenberg interaction $ \vec{d}_j \cdot \vec{d}_{j+1}$, there is no higher-order interaction such as biquadratic $(\vec{d}_j \cdot \vec{d}_{j+1})^2$.
Action of ${\cal H}_{j,j+1}$ on the Hilbert space spanned by $\vert +\rangle_j \,\vert +      \rangle_{j+1}$, $\vert +\rangle_j \,\vert -      \rangle_{j+1}$, $\vert -\rangle_j \,\vert +      \rangle_{j+1}$, and  $\vert -\rangle_j \,\vert -      \rangle_{j+1}$ will also project into the states of $\vec{D}$ and $\vec{Q}$ operators, but those may be ignored for low-lying Hilbert space as they belong to much higher energies. There will be no second nearest neighbor interaction (see end matter) such as $\vec{d}_{j-1} \cdot \vec{d}_{j+1}$. 
The coefficient of $\vec{d}_j \cdot \vec{d}_{j+1}$ is also found as $4J_2/9$ in Ref.~\onlinecite{Cheng2022} following Kadanoff's renormalization approach \cite{Kadanoff1966,Drell1977,Jullien1978,Martin1996,Kargarian2008}. While this procedure treats virtual transition to higher energy Hilbert space of middle trimer for calculating effective next-nearest neighbor interaction, which is very small anyway, it ignores most important degenerate low-energy Hilbert space of the middle trimer. Our explicit calculation (see end matter) shows that for the present case, the latter is, however, identically zero. Therefore, the effective Hamiltonian \eqref{H_effective} is proved to be almost exact for low-energy Hilbert space.

Bethe Ansatz solution of ${\cal H}_{\rm eff}$ provides an exact ground state energy per spin as $E_{\rm gs}^{\rm BA} =-\frac{J_1}{3}[1 - (\gamma/9)(1-4\ln 2)]$ because the ground state energy per spin in a Heisenberg antiferromagnetic chain is $(1/4-\ln 2)$ in the unit of nearest neighbor exchange coupling. The excited states correspond to spinon-continuum bounded by the dispersions
\begin{eqnarray}
	\omega_l &=& \frac{2\pi}{9} J_2 \vert \sin (3q) \vert  \label{dispersion_lower} \\
	\omega_u &=& \frac{4\pi}{9} J_2 \vert \sin (3q/2) \vert \label{dispersion_higher}
\end{eqnarray} 
whose periodicity is one-third of the same for Heisenberg spin-1/2 chain.
We note that these spinons are distinctly different from the usual spinons which occur due to flipping of lattice spins within the domain walls; a similar arrangements of the composite spins of trimers describe the spinons here (Fig.~\ref{schematic}b). This exhausts low-energy Hilbert space of ${\cal H}$ when $\gamma < 9/(4\pi)$. This upper bound of $\gamma$ is obtained by making the energy difference between two lowest energy levels of a single trimer, $\epsilon_1 -\epsilon_0$, greater than the maximum of low-energy excitations, {\it i.e.}, $(4\pi/9) J_2$. Therefore, for $\gamma \lesssim 9/(4\pi)$, the excitation spectrum will have gap beyond energy $(4\pi/9)J_2$. To understand the higher energy excitations, we next perform Jordan-Wigner transformation on ${\cal H}$ followed by its mean-field solution that is benchmarked with the above low-lying excitations.

In Jordan-Wigner transformation\cite{Jordan1928}, conversion of Spin operators into a string of fermion operators obeying anti-commutation relations $\{ c_{j,\lambda},c_{j',\lambda'}^\dagger\} = \delta_{jj'}\delta_{\lambda\lambda'}$, $\{ c_{j,\lambda},c_{j',\lambda'}\} = \{ c_{j,\lambda}^\dagger,c_{j',\lambda'}^\dagger\} =0$
as follows: $S_{j,\lambda}^{+} = S_{j,\lambda}^x + i S_{j,\lambda}^y = c_{j,\lambda}^{\dagger}U_{j,\lambda}^{\dagger}$, $S_{j,\lambda}^{-} = S_{j,\lambda}^x - i S_{j,\lambda}^y = U_{j,\lambda}c_{j,\lambda}$ and $S_{j,\lambda}^{z} = c_{j,\lambda}^{\dagger}c_{j,\lambda}-\frac{1}{2}$ where, $U_{j,\lambda} = \exp\Big[i \pi \sum_{l = 1}^{j-1}\sum_{\lambda'=1}^3 c_{l,\lambda'}^{\dagger}c_{l,\lambda'} \Big(\delta
_{\lambda ,1} + i\pi c_{j,1}^\dagger c_{j,1} (\delta_{\lambda,2} +i\pi c_{j,2}^\dagger c_{j,2} \delta_{\lambda, 3} )\Big)\Big]$. We express the Hamiltonian ${\cal H}$ in terms of fermion operators, followed by the Fourier transformation as $c_{j,\lambda} = \frac{1}{\sqrt{N}}\displaystyle\sum_{k} a_{k,\lambda}e^{-i k j}$, with the normalization $\sum_{j=1}^N e^{i(k-k')j} = N\delta_{kk'}$, and fermionic anti-commutation relations $\{ a_{k,\lambda},a_{k',\lambda'}^\dagger\} = \delta_{kk'}\delta_{\lambda\lambda'}$, $\{ a_{k,\lambda},a_{k',\lambda'}\} = \{ a_{k,\lambda}^\dagger,a_{k',\lambda'}^\dagger\} =0$. We obtain ${\cal H} = {\cal H}_0 + \hat{V}$, where ${\cal H}_0$ consists of terms with bilinear fermionic operators which may easily be diagonalized and constant energy terms, and the operator $\hat{V}$ consists of the terms only with four fermion operators. We consider a mean field approximation (see end matter) in the  Hartree-Fock channel in which $\langle a_{k_1,\lambda_1}^\dagger a_{k_1,\lambda_2} \rangle = \Delta_{\lambda_1,\lambda_2}^{k_1} \neq 0$, where $\langle {\cal O} \rangle$ represents the expectation value of the operator ${\cal O}$ in the ground state, which needs to be determined. However, the mean-field contribution $\langle a_{k_1,\lambda_1}^\dagger a_{k_2,\lambda_2} \rangle$, $(k_1 \neq k_2)$, is ignored. 
Additionally, the BCS channel is closed as $\langle a_{k_1,\lambda_1}^\dagger a_{-k_1,\lambda_2}^\dagger\rangle =0$ for $\lambda_1 \neq \lambda_2$. In this approximation, we find ${\cal H} \approx {\cal H}_{\rm eff} ={\cal H}_0 + \hat{V}_{\rm MF}$, where $\hat{V}_{MF}$ consists of the constant energy terms and the terms with bilinear fermionic operators. 

\begin{figure}[h]	
	\centering   
	\includegraphics[width = \linewidth]{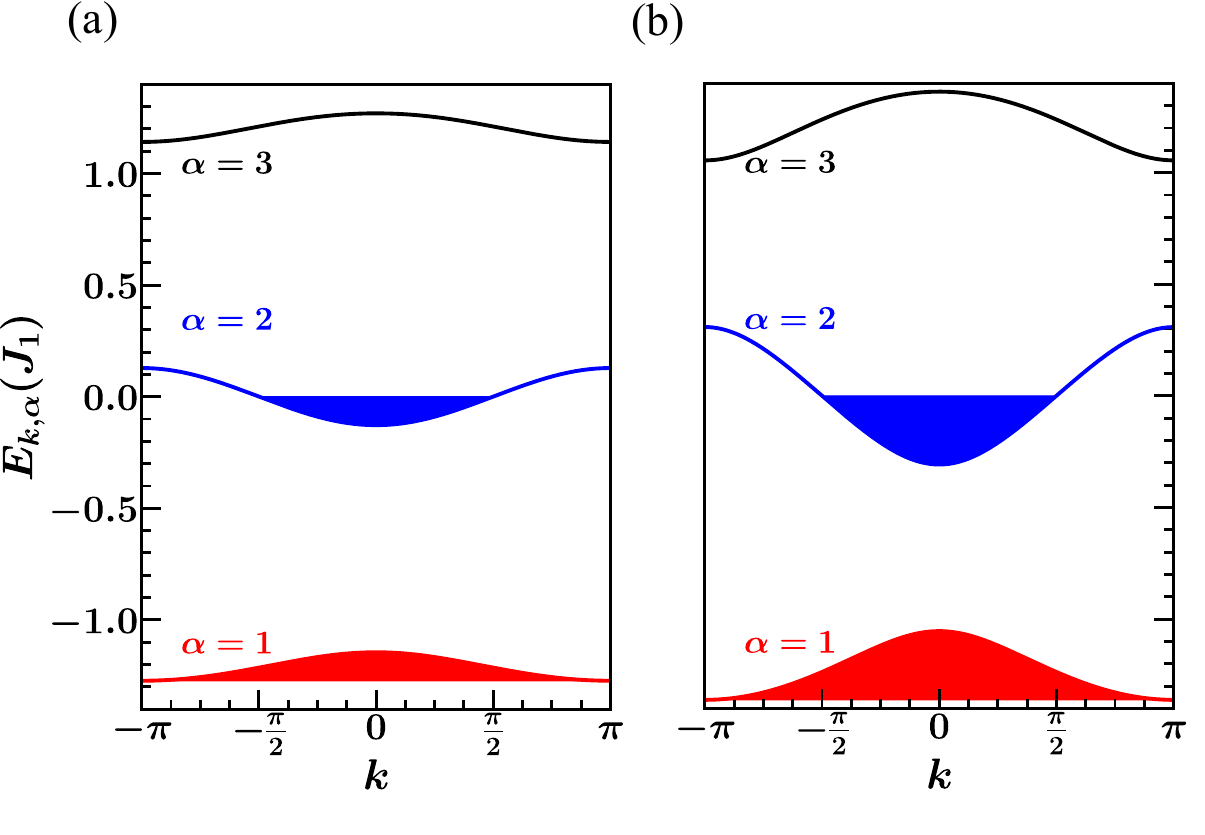}
	\caption{Energy dispersion of three bands at (a) $\gamma =0.2$ and (b) $\gamma = 0.5$. Shaded portions of the bands are filled by fermions in the ground state.}
	\label{energy_band}
\end{figure}

We self-consistently diagonalize ${\cal H}_{\rm eff}$ to obtain
\begin{equation}
	{\cal H}_{\rm eff} = \sum_{k,\alpha} E_{k,\alpha} {\cal A}_{k,\alpha}^\dagger {\cal A}_{k,\alpha} + E_0
\end{equation} 
where renormalized fermionic operators in the diagonal basis, ${\cal A}_{k,\alpha} = \sum_{\lambda=1}^3 b_{\alpha\lambda,k} a_{k,\lambda}$ with the coefficients $b_{\alpha\lambda,k}$ obtained from eigenstates for the eigenvalue $E_{k,\alpha}$ ($\alpha =1,2,3$). The orthonormality conditions of the eigenstates, $\sum_{\lambda}  b_{\alpha\lambda,k} b_{\alpha'\lambda,k} =\delta_{\alpha\alpha'}$ ensures that ${\cal A}_{k,\alpha}$ are linearly independent fermionic operators. In Fig.~\ref{energy_band} , we show all the three bands formed by $E_{k,\alpha}$. All the negative energy states are filled by fermions, {\it i.e.}, the band with $\alpha =1$ is completely filled, and the band with $\alpha=2$ is half-filled, filled for $\vert k \vert \leq \pi/2$. This suggests that the ground state $\Psi_{\rm gs} = \prod_{\vert k \vert \leq  \pi} {\cal A}_{k,1}^\dagger \prod_{\vert k \vert \leq  \pi/2} {\cal A}_{k,2}^\dagger \vert 0\rangle$ corresponds to the ground-state energy per trimer (3 spins)
\begin{equation}
E_{\rm gs} = E_0 + \frac{1}{2\pi}\int_{-\pi}^\pi E_{k,1} \, dk + \frac{1}{2\pi}\int_{-\pi/2}^{\pi/2} E_{k,2}\, dk
\label{energy_gs}
\end{equation} 
and magnetization $M=0$, because $\alpha =1$ band is completely filled, $\alpha =2$ is half-filled, and $\alpha =3$ is completely empty and $z$-component of total spin operator, $ \sum_{j,\lambda} S_{j,\lambda}^z = \sum_{k,\lambda} {\cal A}_{k,\alpha}^\dagger {\cal A}_{k,\alpha} - 3N/2$ in terms of fermionic operators. The expression of energy $E_0$ in Eq.\eqref{energy_gs} is shown in the end matter. For $\gamma = 0.2$ and $0.5$, $E_{\rm gs}$ per spin are found respectively as $-0.334J_1$ and $-0.359J_1$, which are within $2$--$4\%$ of the estimated values of Bethe Ansatz ground state energy $E_{\rm gs}^{\rm BA}$.

\begin{figure}[h]
	\centering   
	\includegraphics[width = \linewidth]{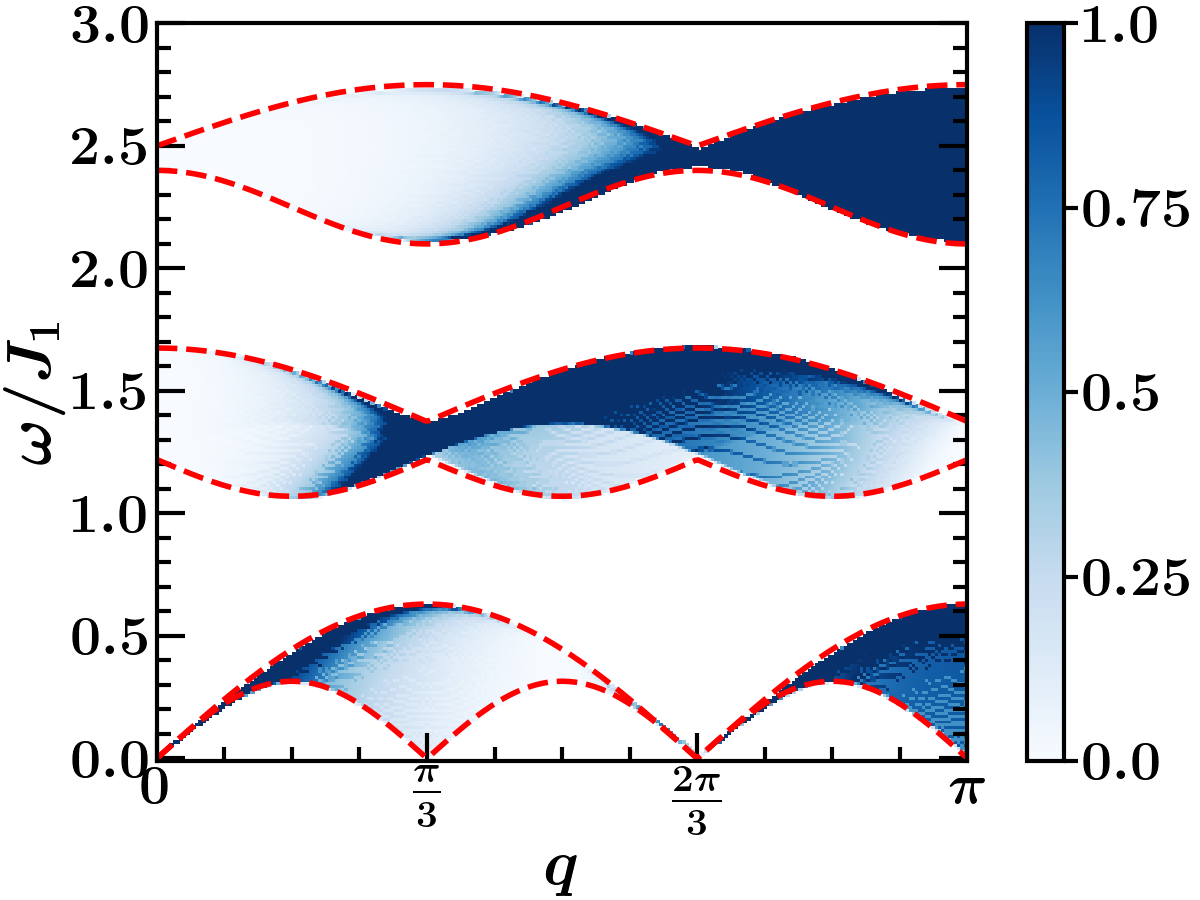}    
	\caption{Dynamical spin structure factor ${\cal P}(\omega,q)$ calculated at $\gamma=0.5$ and zero magnetic field by color coding in $0$--$1$ scale. The red dashed lines have been drawn as a visual guide to the lower-bound and higher-bound of the band. The functional forms of these lines have been quoted in the text.}
	\label{structure_factor}
\end{figure}

The excitation spectrum can be obtained by calculating the dynamical spin structure factor
\begin{equation}
{\cal P} (\omega , q) = \sum_{m} \Big\vert \langle  \Psi_m^{\rm ex} \vert {\cal S}^z (q) \vert \Psi_{\rm gs} \rangle\Big\vert^2 \delta (\omega - E_m^{\rm ex} + E_{\rm gs}) 	
\end{equation}
where 
${\cal S}^z(q) = \frac{1}{\sqrt{3N}} \sum_{j=1}^{N}\sum_{\lambda=1}^3 e^{\imath q(3(j-1)+\lambda)}  S_{j,\lambda}^z $ is the Fourier-transform of $S_{j,\lambda}^z$ and $m$ labels all possible excited states defined by ${\cal A}_{k+3q,\alpha'}^\dagger  {\cal A}_{k,\alpha} \vert \Psi_{\rm gs} \rangle $ with $\alpha$ ($\alpha'$) being occupied (unoccupied) in the ground state. In Fig.~\ref{structure_factor}, we show ${\cal P}(\omega,q)$ for $\gamma = 0.5$ as color-gradient in the $\omega$--$q$ plane. The excitations are found in the form of three bands, as reported earlier in numerical simulations\cite{Bera2022,Cheng2022}. The lowest band is for the excitations with $\alpha=\alpha'=2$, {\i.e.,} within the half-filled band (Fig.~\ref{structure_factor}), the middle band is for two types of excitations: (i) $\alpha =2$ and $\alpha'=3$, and (ii) $\alpha=1$ and $\alpha'=2$. The top-most band is for the excitations defined by $\alpha =1$ and $\alpha'=3$. The lowest energy band is bounded by the dispersions $\omega_{l,1} = 0.32J_1 \vert \sin 3q \vert$ and $\omega_{u,1} = 0.64 J_1\vert \sin (3q/2)\vert$ which are nearly equal to the expressions \eqref{dispersion_lower} and \eqref{dispersion_higher} respectively obtained from the exact low-energy analysis. This remarkable success of the mean-field theory encourages us to predict the bounds of higher energy bands of excitations. The middle band is bounded by the dispersions $\omega_{l,2} = 1.22J_1 -0.15J_1 \vert \sin 3q\vert$ and $\omega_{u,2} = 1.375J_1+ 0.3J_1\vert \cos(3q/2) \vert$.  The top-most excitation band is bounded by the dispersions
$\omega_{l,3} = 2.25J_1 + 0.15J_1 \cos(3q)$ and $\omega_{u,3} = 2.5J_1 + 0.25J_1 \vert \sin (3q/2)\vert$.
These two higher energy excitation bands may be associated with exotic doublon and quarton exciations \cite{Cheng2022,Cheng2024,Bera2022,Li2025} 

\begin{figure}[h]
	\centering
	\includegraphics[width = 0.9\linewidth]{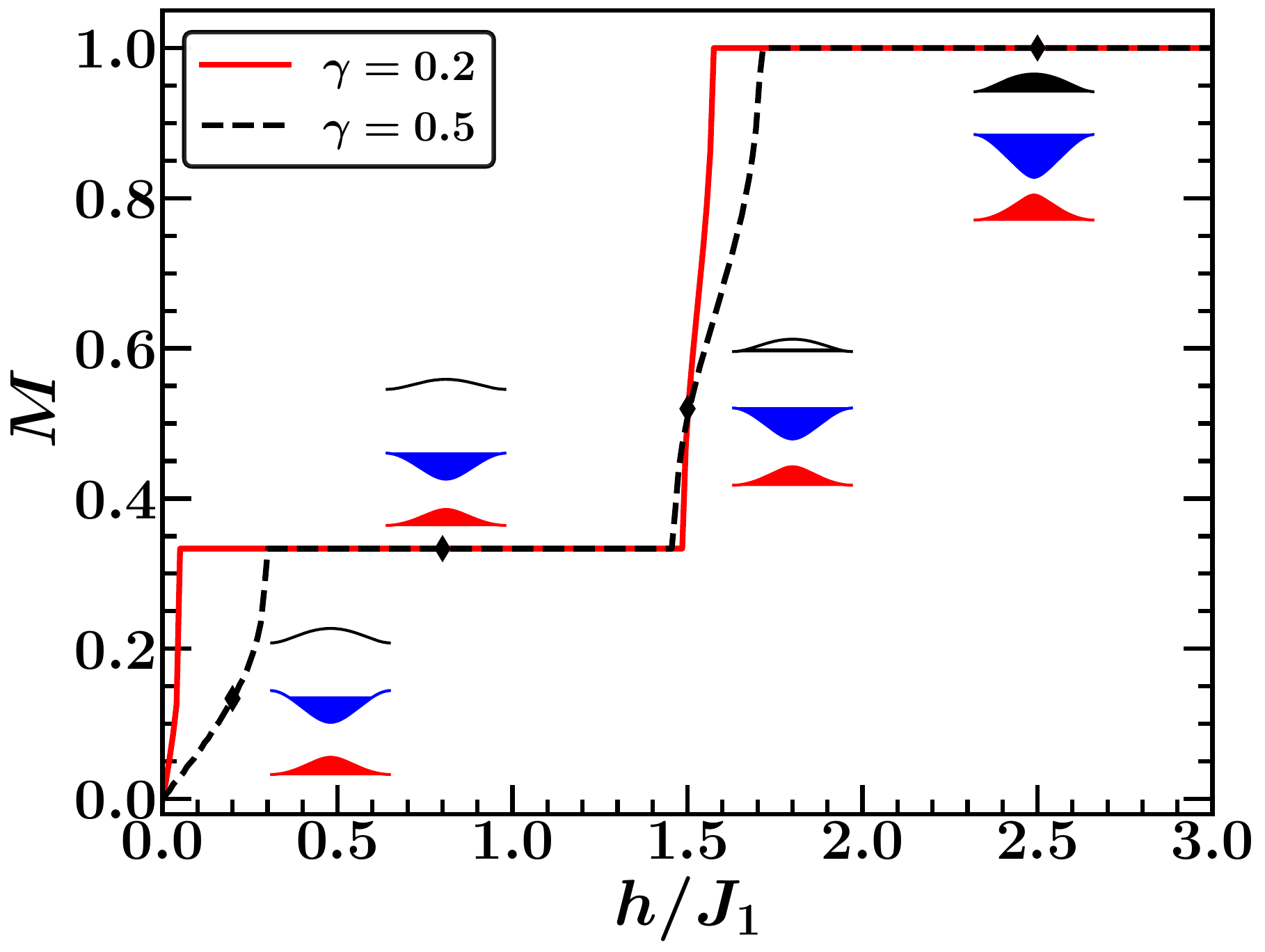}
	\caption{Magnetization per spin $M$ in the unit of saturated magnetization $M_0$ {\it versus} $h$ for $\gamma = 0.2$ (solid line) and $0.5$ (dashed line) at zero temperature. A magnetization plateau occurs at $M=1/3$, apart from the plateau of saturated magnetization. Four regimes $M < 1/3$, $M=1/3$, $1/3 < M <1$ and $M=1$ are depicted by the designated four points to which the change in the filling of bands by fermions are described in the corresponding ground states.}
	\label{magnetization}
\end{figure}

For the application of external magnetic field $H$, an additional term $-h\sum_{j,\lambda} S^z_{j,\lambda}$ with $h= \mu_0g\mu_{_B}H$ needs to be included in the Hamiltonian \eqref{full_hamiltonian}, where $\mu_0$, $g$, and $\mu_{_B}$ respectively denote permeability in free space, $g$-factor, and Bohr-magneton. The mean-field solutions are obtained for different values of $h$. As we increase $h$ from zero, the band with $\alpha =2$ is getting more and more filled and thereby a continuous increase of $M$ from zero (see Fig.~\ref{magnetization}). At a critical field $h_1$, this band becomes completely filled. For $h_1 < h <h_2$, no more filling of bands occurs because of the presence of band gap between the bands with $\alpha =2$ and 3. For $h_2 \leq h \leq h_3$, the band with $\alpha =3$ gets gradually filled, and for $h >h_3$, no more states available to fill. Therefore, in the regimes $h \leq h_1$ and $h_2 \leq h \leq h_3$, M increases with $h$. In the intermediate range of magnetic field $h_1 <h<h_2$, magnetization forms a plateau with $M=1/3$, as observed in the NCGO system, because two out of three bands are completely filled, so that $M = (2N-3N/2)/(3N/2) = 1/3$. The width of the plateau is found to be consistent with the results of numerical simulation \cite{Bera2022}. The magnetization saturates at $M=1$ for $h>h_3$. The 1/3 magnetization plateau is expected in this model because of the presence of 3 sub-lattices for spin-1/2 as $3\times (1/2)(1-1/3)$ is an integer\cite{Oshikawa1997}. 

\begin{figure}[h]
	\centering     
	\includegraphics[width = \linewidth]{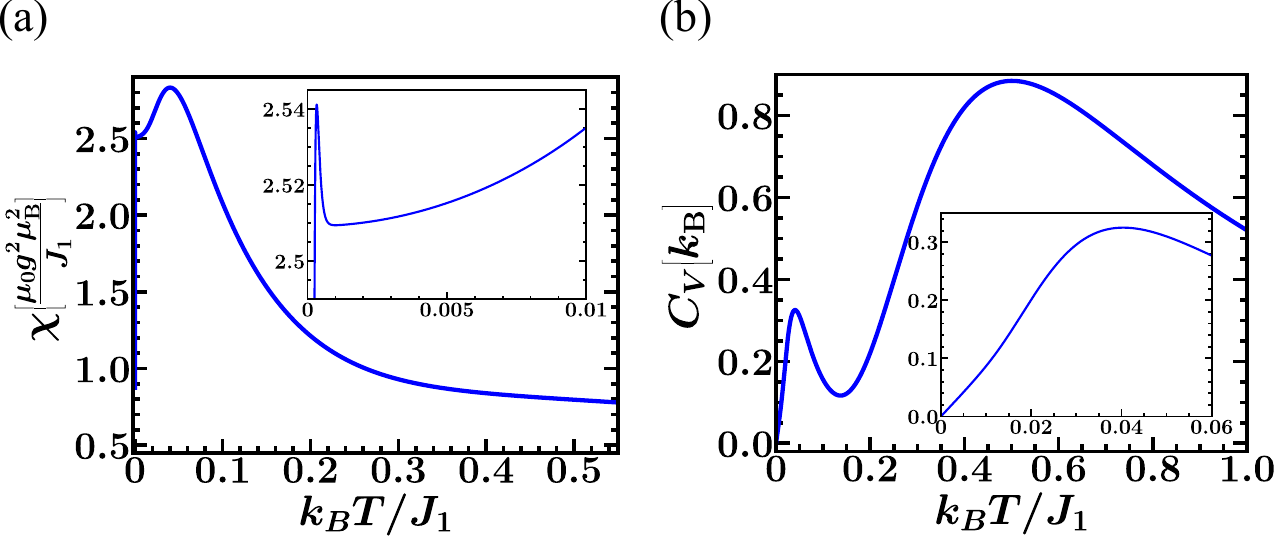}
	\caption{Temperature dependence of (a)Magnetic susceptibility and (b) specific heat. Insets show the corresponding zoomed-view at low temperatures.}
	\label{chi_cv}
\end{figure}

In Fig.~\ref{chi_cv} we show two thermodynamic properties, namely, magnetic susceptibility at zero magnetic field and specific heat, which we calculate using the energy bands (Fig.~\ref{energy_band}) for $\gamma = 0.2$: The magnetic susceptibility per trimer is $\chi = \frac{\mu_0}{k_BT}g^2\mu_B^2 \sum_{\alpha=1}^3\int_{-\pi}^\pi  \frac{dk}{2\pi} f_{k,\alpha}(1-f_{k,\alpha})$ with Fermi function $f_{k,\alpha} = 1/[1+e^{E_{k,\alpha}/k_BT}] $, and the specific heat per trimer is given by $C_v = \frac{1}{k_BT^2} \sum_{\alpha=1}^3\int_{-\pi}^\pi  \frac{dk}{2\pi} E_{k,\alpha}^2 f_{k,\alpha}(1-f_{k,\alpha})$. Recall \cite{Eggert1994,Johnston2000} that $\chi$ acquires a peak at about $k_BT = 0.641 J$ which is onset of low-temperature antiferromagnetic correlation, and an inflection point at about $k_BT= 0.087J$ below which slope of $\chi \to \infty$ as $T\to 0$ in a spin-1/2 antiferromagnetic chain with the coupling strength $J$. Since the low-energy effective Hamiltonian is nothing but an effective Heisenberg interaction \eqref{H_effective}, one would then expect similar phenomena for the trimer chain, respectively at $k_BT =0.057 J_1$ and $0.008 J_1$ that are consistent to the temperatures found by our explicit calculation (see Fig.\ref{chi_cv}a), except that the inflection point appears as a minor peak that is resolved by increasing the temperature resolution. Similarly, the specific heat $C_v$ is expected \cite{Johnston2000} to have a peak at $k_BT = 0.044J_1$ that also agrees with our finding (see Fig.\ref{chi_cv}b). Therefore, the thermodynamic properties of the trimer chain at low temperatures further confirms the emergent low-energy effective Heisenberg Hamiltonian \eqref{H_effective}. In addition to this, higher energy excitations lead to increase in $C_v$ with temperature until it reaches to a maximum value at $k_BT = 0.5J_1$ followed by its decrease as $T^{-2}$ at higher temperatures.

We show an emergent effective Hamiltonian for low-energy dynamics, and have performed a mean-field treatment of the entire Hamiltonian. It is astounding that the latter almost exactly reproduces the former, and additionally predicts higher-energy dynamics. Some of the predictions are consistent with experiments and numerical simulations, and the rest warrant experimental verifications. The trimerized spin model with uniform interaction between intra-trimer spins that have been considered here is relevant for  Na$_2$Cu$_3$Ge$_4$O$_{12}$ compound.  A similar treatment may also be employed for another trimerized chain\cite{Verkholyak2021} where interactions between intra-trimer spins are unequal, relevant for Cu$_3$(P$_2$O$_6$OH)$_2$  compound\cite{Hase2006}. In this case, however, low-energy Hilbert space will not be formed by the spins $\vec{d}_j$ but by interaction-dependent linear combinations of $\vec{d}_j$ and $\vec{D}_j$. This intricacy, along with the context of other experiments, such as Raman scattering and resonant inelastic x-ray scattering will be published elsewhere\cite{Note}. Our analysis may further be generalized to two-dimensional trimerized spin systems for obtaining an effective Hamiltonian for predicting low-energy dynamics and excitations.

Acknowledgments: We thank Anushree Roy and S. M. Yusuf for introducing to the important experimental data of spin-trimer systems, and Diptiman Sen and Subhro Bhattacharjee for enlightening discussions. SSM acknowledges support from ANRF grant No ANRF/ARGM/2025/000503/TS.

\bibliographystyle{apsrev4-2}

%

\section{End Matter}

\noindent {\bf Next-nearest neighbor inter-trimer interaction:}

In the lowest-energy basis of three consecutive trimers, given by  $\vert +\rangle_j \vert +  \rangle_{j+1} \vert + \rangle_{j+2}$, $\vert +\rangle_j \vert +  \rangle_{j+1} \vert - \rangle_{j+2}$, $\vert +\rangle_j \vert -  \rangle_{j+1} \vert + \rangle_{j+2}$, $\vert +\rangle_j \vert -  \rangle_{j+1} \vert - \rangle_{j+2}$, $\vert -\rangle_j \vert +  \rangle_{j+1} \vert + \rangle_{j+2}$, $\vert -\rangle_j \vert +  \rangle_{j+1}\, \vert - \rangle_{j+2}$, $\vert -\rangle_j \vert -  \rangle_{j+1} \vert + \rangle_{j+2}$, and $\vert -\rangle_j \vert -  \rangle_{j+1} \vert - \rangle_{j+2}$, the inter-trimer interacting Hamiltonian ${\cal H}_I={\cal H}_{j,j+1}+{\cal H}_{j+1,j+2}$ becomes
\begin{equation}
	J_2\left(
	\begin{array}{rrrrrrrr}
		\frac{2}{9} & 0 & 0 & 0 & 0 & 0 & 0 & 0 \\
		0 & 0 & \frac{2}{9} & 0 & 0 & 0 & 0 & 0 \\
		0 & \frac{2}{9} & -\frac{2}{9} & 0 & \frac{2}{9} & 0 & 0 & 0 \\
		0 & 0 & 0 & 0 & 0 & \frac{2}{9} & 0 & 0 \\
		0 & 0 & \frac{2}{9} & 0 & 0 & 0 & 0 & 0 \\
		0 & 0 & 0 & \frac{2}{9} & 0 & -\frac{2}{9} & \frac{2}{9} & 0 \\
		0 & 0 & 0 & 0 & 0 & \frac{2}{9} & 0 & 0 \\
		0 & 0 & 0 & 0 & 0 & 0 & 0 & \frac{2}{9} \\
	\end{array}
	\right) \, .
\end{equation}
The interaction matrix for two consecutive trimers, namely, $j$-th and $(j+1)$-th trimers may be found using this matrix by considering matrix elements of ${\cal H}_I$ in the basis vectors $\frac{1}{\sqrt{2}}\Big( 
\vert \sigma_1\rangle_j \vert \sigma_2 \rangle_{j+1} \vert + \rangle_{j+2} + \vert \sigma_1\rangle_j \vert \sigma_2 \rangle_{j+1} \vert - \rangle_{j+2}\Big)$ where $\sigma_{1,2} = \pm$. The corresponding matrix is exactly same as found in Eq.\eqref{matrix-4} by considering the matrix elements between two consecutive trimers only for obtaining nearest-neighbor inter-trimer interaction in low-energy Hilbert space. Similarly, the next-nearest neighbor interaction between trimers, namely, $j$-th and $(j+2)$-th trimers may be obtained by calculating the matrix elements of ${\cal H}_I$ in the basis vectors $\frac{1}{\sqrt{2}}\Big( 
\vert \sigma_1\rangle_j \vert +\rangle_{j+1} \vert \sigma_2 \rangle_{j+2} + \vert \sigma_1\rangle_j \vert - \rangle_{j+1} \vert \sigma_2\rangle_{j+2}\Big)$. We find that all the matrix elements in this case are identically zero. In other words, {\em no} interaction is involved between next-nearest neighbor trimers.

\begin{widetext}
	
\noindent {\bf Mean Field:} Performing Jordan-Wigner transformation, we find ${\cal H}$ in Eq.\eqref{full_hamiltonian} as		
  				\begin{eqnarray}\label{J-W_Hamiltonian}
  					{\cal H} &=& \sum_{j = 1}^{N}\Bigg[ J_1\sum_{\lambda = 1}^{2}\Bigg\{c_{j,\lambda}^{\dagger} c_{j,\lambda} c_{j,\lambda+1}^{\dagger}c_{j,\lambda+1}+\frac{1}{2}\Big(c_{j,\lambda}^{\dagger}c_{j,\lambda+1}+\text{h.c.}\Big) -\frac{1}{2}\Big(c_{j,\lambda}^{\dagger}c_{j,\lambda}+c_{j,\lambda+1}^{\dagger}c_{j,\lambda+1}\Big) \Bigg\} 
  					+J_2\Bigl\{c_{j,3}^{\dagger}c_{j,3}c_{j+1,1}^{\dagger}c_{j+1,1} \nonumber \\
  					&&+\frac{1}{2}\Big(c_{j,3}^{\dagger}c_{j+1,1}+\text{h.c.}\Big)  -\frac{1}{2}\Big(c_{j,3}^{\dagger}c_{j,3}+c_{j+1,1}^{\dagger}c_{j+1,1}\Big)\Bigl\}  +\Big(\frac{J_1}{2}+\frac{J_2}{4}\Big)-h\sum_{\lambda=1}^3\Big(c_{j,\lambda}^\dagger c_{j,\lambda} - \frac{1}{2} \Big) \Bigg]	
  					\end{eqnarray} 
  when Zeeman coupling $-h\sum_{j,\lambda} S^z_{j,\lambda}$ is also added to the Hamiltonian ${\cal H}$ in  Eq.~\ref{full_hamiltonian} due to the application of external magnetic field.				

Performing Fourier transform of the fermionic operators, we obtain ${\cal H} = {\cal H}_0 +\hat{V}$, where
  \begin{eqnarray}\label{full_hamiltonian_fourier_first}
  	{\cal H}_{0}&=&\sum_{k}\Bigg[J_1\sum_{\lambda = 1}^{2} \Big[-\frac{1}{2}\Big(a_{k,\lambda}^{\dagger}a_{k,\lambda}+a_{k,\lambda+1}^{\dagger}a_{k,\lambda+1}\Big) 
  	+\frac{1}{2}\Big(a_{k,\lambda}^{\dagger}
  	a_{k,\lambda+1}+a_{k,\lambda+1}^{\dagger}a_{k,\lambda}\Big)\Big]+J_2\Big[-\frac{1}{2}\Big(a_{k,1}^{\dagger}a_{k,1} 
  	 +a_{k,3}^{\dagger}
  	a_{k,3}\Big) \nonumber \\
  	&&+\frac{1}{2}\Big(e^{ik}a_{k,1}^{\dagger}a_{k,3}+e^{-ik}a_{k,3}^{\dagger}a_{k,1}\Big)\Big] 
  	  -h\sum_{\lambda =1}^3 a_{k,\lambda}^\dagger a_{k,\lambda}   \Bigg] + N\Big( \frac{J_1}{2}+\frac{J_2}{4} + \frac{3h}{2}\Big)
  \end{eqnarray}
  by considering lattice constant to be unity, and 
  \begin{eqnarray}\label{full_hamiltonian_fourier_second}
  	\hat{V} &=&\frac{1}{N}\sum_{k_1, k_2}\Bigg[J_1\sum_{\lambda = 1}^{2}\Big(a_{k_1,\lambda}^{\dagger}a_{k_1,\lambda}a_{k_2,\lambda+1}^{\dagger}a_{k_2,\lambda+1}  -a_{k_1,\lambda}^{\dagger}a_{k_1,\lambda+1}a_{k_2,\lambda+1}^{\dagger}a_{k_2,\lambda}\Big)\nonumber \\ &&
  	+J_2\Big(a_{k_1,3}^{\dagger}a_{k_1,3}a_{k_2,1}^{\dagger}a_{k_2,1} -a_{k_1,3}^{\dagger}a_{k_1,1} a_{k_2,1}^{\dagger}a_{k_2,3}\,e^{i\left(k_2-k_1\right)}\Big)\Bigg] \, ,
  \end{eqnarray}
where the first(second) term within the parentheses represent direct(exchange) interaction.
While ${\cal H}_0$ is exactly diagonalizable, $\hat{V}$ needs to be approximated to express in terms of bilinear fermionic operators so that the resultant Hamiltonian becomes diagonalizable.
In the mean-field approximation as described in the main text, $\hat{V}$ becomes
{\small \begin{eqnarray}
 && \hat{V}_{\rm MF} =\frac{1}{N} \sum_{k_1,k_2} \Bigg[ J_1 \sum_{\lambda=1}^2 \Big( \Delta_{\lambda+1,\lambda+1}^{k_2}\, a_{k_1,\lambda}^{\dagger}a_{k_1,\lambda} + \Delta_{\lambda,\lambda}^{k_1}   a_{k_2,\lambda+1}^{\dagger}a_{k_2,\lambda+1} - \Delta_{\lambda,\lambda}^{k_1}\Delta_{\lambda+1,\lambda+1}^{k_2}   -\Delta_{\lambda+1,\lambda}^{k_2}\,a_{k_1,\lambda}^{\dagger}a_{k_1,\lambda+1}  - \Delta_{\lambda,\lambda+1}^{k_1}\,  a_{k_2,\lambda+1}^{\dagger}a_{k_2,\lambda}\nonumber \\ &&+  \Delta_{\lambda,\lambda+1}^{k_1}\Delta_{\lambda+1,\lambda}^{k_2} \Big)+J_2\Big( \Delta_{3,3}^{k_1}\,  a_{k_2,1}^{\dagger}a_{k_2,1}  +\Delta_{1,1}^{k_2}\, a_{k_1,3}^{\dagger}a_{k_1,3}  - \Delta_{3,3}^{k_1} \Delta_{1,1}^{k_2}  
  - e^{i\left(k_2-k_1\right)}  \Big\{ \Delta_{3,1}^{k_1}  a_{k_2,1}^{\dagger}a_{k_2,3}  + \Delta_{1,3}^{k_2}a_{k_1,3}^{\dagger}a_{k_1,1}    - \Delta_{3,1}^{k_1} \Delta_{1,3}^{k_2}\Big\} \,\Big)\Bigg] \nonumber \\
\end{eqnarray}
}
We next express ${\cal H}_{\rm eff} = {\cal H}_0 + \hat{V}_{\rm MF}$  in the basis $\Psi_k^T \equiv (a_{k,1},\, a_{k,2},\,a_{k,3})^T$ as
$	{\cal H}_{\rm eff} = \sum_k \Psi_k^T {\cal M}_k \Psi_k + E_0 $
where the matrix ${\cal M}_k$ is given by
{\small
\begin{eqnarray}
	{\cal M}_k =  \left(\begin{array}{ccc}
	-\frac{(J_1+J_2+2h)}{2}+\frac{1}{N}\sum_{k'}(J_1\Delta_{2,2}^{k'}+J_2\Delta_{3,3}^{k'}) &\frac{J_1}{2} - \frac{J_1}{N}\sum_{k'}\Delta_{2,1}^{k'} & \frac{J_2}{2}e^{ik} - \frac{J_2}{N}\sum_{k'}\Delta_{3,1}^{k'}e^{i(k-k')} \\  & & \\
	\frac{J_1}{2} - \frac{J_1}{N}\sum_{k'}\Delta_{1,2}^{k'} & -(J_1+h) +\frac{J_1}{N}\sum_{k'}(\Delta_{1,1}^{k'}+ \Delta_{3,3}^{k'}) & \frac{J_1}{2} - \frac{J_1}{N}\sum_{k'}\Delta_{3,2}^{k'} \\ & & \\
	\frac{J_2}{2}e^{-ik} - \frac{J_2}{N}\sum_{k'}\Delta_{1,3}^{k'}e^{-i(k-k')} \,\,\,\,\, & \frac{J_1}{2} - \frac{J_1}{N}\sum_{k'}\Delta_{2,3}^{k'}  & -\frac{(J_1+J_2+2h)}{2}+\frac{1}{N}\sum_{k'}(J_1\Delta_{2,2}^{k'} +J_2\Delta_{1,1}^{k'})
 \end{array} \right) \nonumber \\
\label{matrix_equation}
\end{eqnarray}
}
and a constant (independent of fermionic operators) energy
{\small \begin{eqnarray}
	E_0 &=&  N\left(\frac{J_1}{2} + \frac{J_2}{4} + \frac{3h}{2} \right)+ \frac{J_1}{N}\sum_{k,k'} \Big( -\Delta_{1,1}^k\Delta_{2,2}^{k'} - \Delta_{2,2}^k\Delta_{3,3}^{k'} +\Delta_{1,2}^k\Delta_{2,1}^{k'} +\Delta_{2,3}^k\Delta_{3,2}^{k'} 
	 -\gamma \Delta_{3,3}^k\Delta_{1,1}^{k'} +\gamma e^{i(k-k')} \Delta_{1,3}^k\Delta_{3,1}^{k'} \Big) \nonumber \\
\end{eqnarray} }
By diagonalizing ${\cal M}_k$ with eigenvalues $E_{k,\alpha}$ ($\alpha =1,2,3$), one finds
$	{\cal H}_{\rm eff} = \sum_{k,\alpha} E_{k,\alpha} {\cal A}_{k,\alpha}^\dagger {\cal A}_{k,\alpha} + E_0 $
with the operators 
 $ {\cal A}_{k,\alpha} = \sum_{\lambda=1}^3 b_{\alpha\lambda,k} a_{k,\lambda}$
where $b_{\alpha\lambda,k}$ are obtained from eigenstate for the eigenvalue $E_{k,\alpha}$. The orthonormality conditions of the eigenstates $\sum_{\lambda}  b_{\alpha\lambda,k} b_{\alpha'\lambda,k} =\delta_{\alpha\alpha'}$ ensures that ${\cal A}_{k,\alpha}$ are linearly independent fermionic operators. As $a_{k,\lambda} = \sum_\alpha b^{-1}_{\lambda\alpha,k} {\cal A}_{k,\alpha}$, we find
\begin{equation}
	\Delta_{\lambda,\lambda'}^k = 
	\sum_{\alpha} b^{-1 \ast}_{\lambda\alpha,k} b^{-1}_{\lambda'\alpha,k} n_k^\alpha 
	\label{Delta}
\end{equation}
where $\langle {\cal A}_{k,\alpha}^\dagger {\cal A}_{k,\alpha'} \rangle = n_k^\alpha \delta_{\alpha\alpha'}$
with $n_k^\alpha = 1 (0)$ being occupied (unoccupied) fermionic state labeled by $\{k,\alpha\}$, corresponding to negative (positive) value of $E_{k,\alpha}$.
Successively diagonalizing Eq.~\eqref{matrix_equation} and calculating $ \Delta_{\lambda,\lambda'}^k $ in Eq.~\eqref{Delta} and thereby updating $n_k^\alpha$, we self-consistently obtain $E_{k,\alpha}$ and $E_0$. Self-consistent solutions are obtained for different values of $h$.\\

\noindent {\bf Fourier-Transformed Spin Operator:} 
The Fourier-transformed spin-operator is given by
 \begin{equation}
	{\cal S}^z(q) = \frac{1}{\sqrt{3N}} \sum_{j=1}^{N}\sum_{\lambda=1}^3 e^{\imath q(3(j-1)+\lambda)}  S_{j,\lambda}^z 
\end{equation}
which may be expressed in terms of fermionic operators following the Jordan-Wigner transformation as     
\begin{eqnarray}\label{S_q_op}
	\mathcal{S}^z(q) &=& \frac{1}{\sqrt{3N}}\Bigg[ \sum_{k}\Big\{a_{k,1}^\dagger a_{k+3q,1} e^{-i2q}+a_{k,2}^\dagger a_{k+3q,2} e^{-iq}+a_{k,3}^{\dagger}a_{k+3q,3}\Big\}
	-\sum_{j = 1}^{N} e^{i3(j-1)q}\Big(\frac{1}{2} e^{iq}+\frac{1}{2}e^{2iq}+\frac{1}{2}e^{3iq}\Big)\Bigg]
\end{eqnarray}
It may be further expressed in terms of fermionic operators in the diagonalized basis of the mean-field Hamiltonian as follows:
\begin{eqnarray}
	\mathcal{S}^z(q) &=& \frac{1}{\sqrt{3N}}\Bigg[\sum_{k}\sum_{\alpha,\alpha'}\Big(b_{1\alpha,k}^{-1\ast}b_{1\alpha',k+3q}^{-1}e^{-i2q}+b_{2\alpha,k}^{-1\ast}b_{2\alpha',k+3q}^{-1}e^{-iq}+b_{3\alpha,k}^{-1\ast}b_{3\alpha',k+3q}^{-1}\Big)\mathcal{A}_{k,\alpha}^\dagger\mathcal{A}_{k+3q,\alpha'}\Bigg]
\end{eqnarray}
neglecting terms which will not contribute to the dynamical spin structure factor.
Here $\alpha(\alpha')$ denotes the occupied(unoccupied) fermionic state in the ground state.

\end{widetext}

\end{document}